\documentclass{emulateapj}
\usepackage{natbib}
\usepackage{graphicx}
\usepackage{amssymb}
\usepackage{amsmath}
\usepackage{wasysym}
\usepackage{color}
\usepackage[normalem]{ulem}
\usepackage{psfrag}
\usepackage[hang,flushmargin]{footmisc}
\graphicspath{{./}}

\newcommand{\halpha}{\hbox{H$\alpha$}}
\newcommand{\hbeta}{\hbox{H$\beta$}}
\newcommand{\hgamma}{\hbox{H$\gamma$}}

\newcommand{\rthree}{\hbox{$R_3$}}
\newcommand{\rtwothree}{\hbox{$R_{23}$}}
\newcommand{\oaur}{\hbox{[\ion{O}{3}] $\lambda 4363$}}
\newcommand{\fiveoo}{\hbox{[\ion{O}{3}] $\lambda 5007$}}
\newcommand{\oaurratio}{\hbox{[\ion{O}{3}] $\lambda$4363/$\lambda\lambda(4959+5007)$}}
\newcommand{\sulratio}{\hbox{[\ion{S}{3}] $\lambda$6717/$\lambda$6731}}
\newcommand{\thirtyseven}{\hbox{[\ion{O}{2}] $\lambda \lambda$3727,\,29}}
\newcommand{\ttwo}{\hbox{$T_2$}}
\newcommand{\tthree}{\hbox{$T_3$}}
\newcommand{\teoii}{\hbox{$T_e$([\ion{O}{2}])}}
\newcommand{\tenii}{\hbox{$T_e$([\ion{N}{2}])}}

\newcommand{\teoiii}{\hbox{$T_e$([\ion{O}{3}])}}
\newcommand{\teneiii}{\hbox{$T_e$([\ion{Ne}{3}])}}
\newcommand{\teariii}{\hbox{$T_e$([\ion{Ar}{3}])}}

\newcounter{minirefcount}
\begin{document}

\title{Direct Method Gas Phase Oxygen Abundances of 4 Lyman Break Analogs\footnotemark[*]}

\author{Jonathan S. Brown\altaffilmark{1}, Kevin V. Croxall\altaffilmark{1}, Richard W. Pogge\altaffilmark{1,2}}  

\footnotetext[*]{Based on data acquired using the Large Binocular Telescope (LBT). The LBT is an international collaboration among institutions in the United States, Italy, and Germany. LBT Corporation partners are: The University of Arizona on behalf of the Arizona university system; Istituto Nazionale di Astrofisica, Italy; LBT Beteiligungsgesellschaft, Germany, representing the Max-Planck Society, the Astrophysical Institute Potsdam, and Heidelberg University; The Ohio State University, and The Research Corporation, on behalf of The University of Notre Dame, University of Minnesota and University of Virginia}

\altaffiltext{1}{Department of Astronomy, The Ohio State University, Columbus,
  OH 43201, USA}  
\altaffiltext{2}{Center for Cosmology and Astro-Particle Physics, The Ohio State University, Columbus, OH 43201, USA}

\slugcomment{Draft Version of  \textit{\today}}

\begin{abstract}
We measure the gas-phase oxygen abundances in 4 Lyman Break Analogs (LBAs) using auroral emission lines to derive direct abundances. The direct method oxygen abundances of these objects are generally consistent with the empirically-derived strong-line method values, confirming that these objects are low oxygen abundance outliers from the Mass-Metallicity (MZ) relation defined by star forming SDSS galaxies. We find slightly anomalous excitation conditions (Wolf-Rayet features) that could potentially bias the empirical estimates towards high values if caution is not exercised in the selection of the strong-line calibration used.  The high rate of star formation and low oxygen abundance of these objects is consistent with the predictions of the Fundamental Metallicity Relation (FMR), in which the infall of relatively unenriched gas simultaneously triggers an episode of star formation and dilutes ISM of the host galaxy. 
\end{abstract}


\section{Introduction}
\label{sec:intro}
Identifying and quantifying correlations between fundamental parameters gives us insight into the physical processes governing the evolution of the objects under investigation. In the $\Lambda$CDM paradigm, galaxies accrete mass primarily via heirarchical mergers with other galaxies. This formulation reproduces the physical properties of galaxies we observe in the nearby universe \citep{Kauffmann00,Hopkins06}. Thus, the mass of a galaxy reveals the gross characteristics of its history. Similarly, the metallicity of a galaxy is a fundamental characteristic which is intimately related to its formation and subsequent chemical evolution. Studying how the mass and metallicity of galaxies correlate across a wide range of physical parameters informs us about how today's galaxies coalesced and evolved over cosmic time. 

The relation between a galaxy's mass and gas phase oxygen abundance (the MZ relation) was first investigated by \citet{Lequeux79}. Subsequent studies often focused on the more readily measured correlation between luminosity and oxygen abundance \citep[the LZ relation; e.g.,][]{Garnett87,Skillman89,Zaritsky94}. With data from Sloan Digital Sky Survey \citep[SDSS;][]{York00} for a very large number of galaxies, \citet{Tremonti04} showed the MZ relation persists across at least 3 orders of magnitude in mass and an order of magnitude in oxygen abundance. This trend was extended 2.5 orders of magnitude lower in mass and another order of magnitude lower in oxygen abundance by \citet{Lee06}. There have been a number of following studies that have investigated possible variations in the MZ relation as a function of redshift \citep[e.g.][]{Erb06}, star formation rate \citep[e.g.][]{Andrews13}, morphology and environment \citep[e.g.][]{Ellison08Apj,Ellison08Aj}, or a combination of these factors \citep[e.g.][]{Mannucci10,LaraLopez10}.

While general trends between galactic parameters are both interesting and useful, objects that deviate from the observed relations offer a unique perspective, as it is these objects which allow for the direct indentification of important physical mechanisms driving galactic evolution. The ``Lyman Break Analogs'' (LBA) project \citep{Heckman05} identified a class of galaxies that appear to deviate from the local galaxy population and more closely resemble high-redshift Lyman Break Galaxies \citep[LBGs; for a review see][]{Giavalisco02}. These objects were initially identified as nearby ($z < 0.3$), compact ($I_{FUV} > 10^9$~$L_{\odot}$~kpc$^{-2}$), and UV bright ($L_{FUV} > 10^{10.3}$~$L_{\odot}$) objects, mimicking the physical conditions in LBGs seen at much higher redshifts. Thus, if LBAs are true analogs of LBGs, they provide us with an opportunity to study a mode of star formation that may have been dominant in the early universe in a much more detailed way than the very distant LBGs allow. After the identification of LBAs, subsequent work \citep{Hoopes07,Basu-Zych07,Overzier08,Overzier09,Overzier10,Goncalves10} used the Hubble Space Telescope (HST), {\it Spitzer}, VLA, Sloan Digital Sky Survey (SDSS), Galaxy Evolution Explorer (GALEX), and the Keck II telescope to investigate the physical properties of these systems, as well as the degree to which they may resemble LBGs. 

The gas-phase oxygen abundance of these objects was estimated by \citet{Overzier09,Overzier10} using the N2 and O3N2 empirical calibrations from \citet[][hereafter PP04]{Pettini04}. After applying the N2 relation to SDSS galaxies and LBAs, \citet{Overzier10} showed the offset of the LBAs from the MZ relation of local star forming galaxies is inversely correlated with mass, with the least massive LBAs falling $\gtrsim 0.2$ dex below the locus of SDSS galaxies. Low metallicity objects are of particular interest, as they offer a view of how the earliest stars and galaxies formed \citep[e.g.][]{Kunth00,Skillman13}. Fortunately the empirical abundance calibrations incorporate many low mass, low metallicity blue compact galaxies (BCGs) in the hope of extending the calibrations to the lowest metallicities possible. However, we have no {\it a priori} reason to expect that a locally calibrated empirical abundance relation ought to apply to a class of exotic objects experiencing an episode of relatively extreme, concentrated star formation, as is found in the centers of these LBAs. 

Typical LBAs exhibit star formation rates an order of magnitude higher than local dwarf galaxies. Additionally, most LBAs have morphologies and kinematics consistent with recent interactions \citep{Overzier09,Overzier10,Goncalves10}. Clearly these objects depart from the physical parameter space occupied by local \ion{H}{2} regions and dwarf galaxies used to calibrate the empirical relations. Are the empircally estimated oxygen abundances of LBAs being systematically affected by their extreme physical conditions? For instance, \citet{Pilyugin10} showed that many of the line ratios used in the typical strong-line abundance indicators \citep{Pagel79,Alloin79} are complex functions of electron temperature. They show that deriving an abundance calibration from a sample of relatively cool \ion{H}{2} regions and applying it to relatively hot \ion{H}{2} regions could yield erroneous abundance estimates. Alternatively, if locally calibrated empirical relations return reliable abundance estimates when applied to LBAs, this would also be of interest, as this is not immediately obvious given the extreme physical nature of these objects.

Fortunately, we are not required to rely on empirical calibrations alone for these objects. With a measure of the electron temperature ($T_e$), we can determine the oxygen abundance directly using the $T_e$ or ``direct'' method \citep{Dinerstein90}. The electron temperature can be determined from temperature sensitive intensity ratios of collisionally excited forbidden lines. Generally speaking, as metallicity increases, the temperature of the nebula decreases, as there are more ions available to cool the gas. In relatively low metallicity nebulae, a measure of the electron temperature is typically obtained using the $\oaurratio$ line ratio. However, somewhat problematically, the auroral oxygen line $\oaur$\,\AA\ is intrinsically faint, making it notoriously difficult to measure in distant and/or faint objects (though see \citet{Hoyos05,Kakazu07,Amorin10,Amorin12} for instances of the direct method being applied at relatively high redshifts). 

We have measured $\oaur$\,\AA\ in four of the objects from the \citet{Overzier09} sample using the newly commissioned Multi-Object Double Spectrograph \#1 (MODS1) on the 8.4m Large Binocular Telescope (LBT). We use the [\ion{O}{3}] line fluxes to determine an electron temperature, yielding a gas-phase oxygen abundance measurement that we can compare to the values derived via empirical techniques.

In Section~\ref{sec:obsred} we describe the observations and data reduction. Section~\ref{sec:analysis} describes the analysis of the data, including the subtraction of the underlying stellar continuum. In Section~\ref{sec:results} we present the results of our analysis. Lastly, in Section~\ref{sec:discussion} we discuss where LBAs fit in the bigger picture of galaxy formation and evolution; Section~\ref{sec:summary} provides a summary. Throughout this paper we assume $H_0 = 70$ km s$^{-1}$ Mpc$^{-1}$, $\Omega_{\lambda} = 0.7$, and $\Omega_M = 0.3$. With these cosmological parameters, a redshift of $z=0.2$ corresponds to an age of the universe of $\sim 11$ Gyr.

\section{Observations and Reduction}
\label{sec:obsred}

\begin{figure*}
\centering{\includegraphics[scale=1.,width=\textwidth,trim=0.pt 0.pt 0.pt 0.pt,clip]{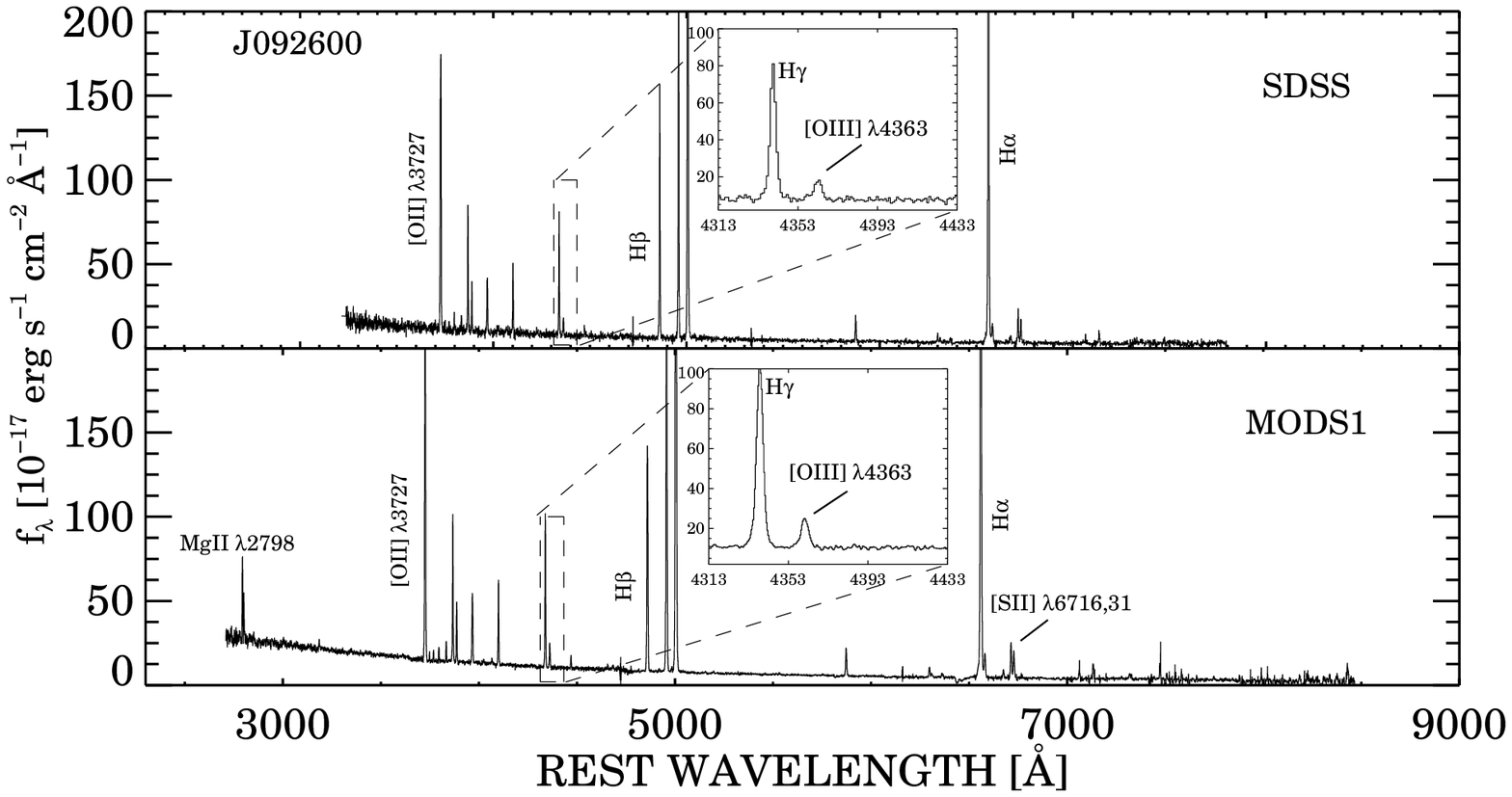}}
\end{figure*}

\begin{figure*}
\centering{\includegraphics[scale=1.,width=\textwidth,trim=0.pt 0.pt 0.pt 0.pt,clip]{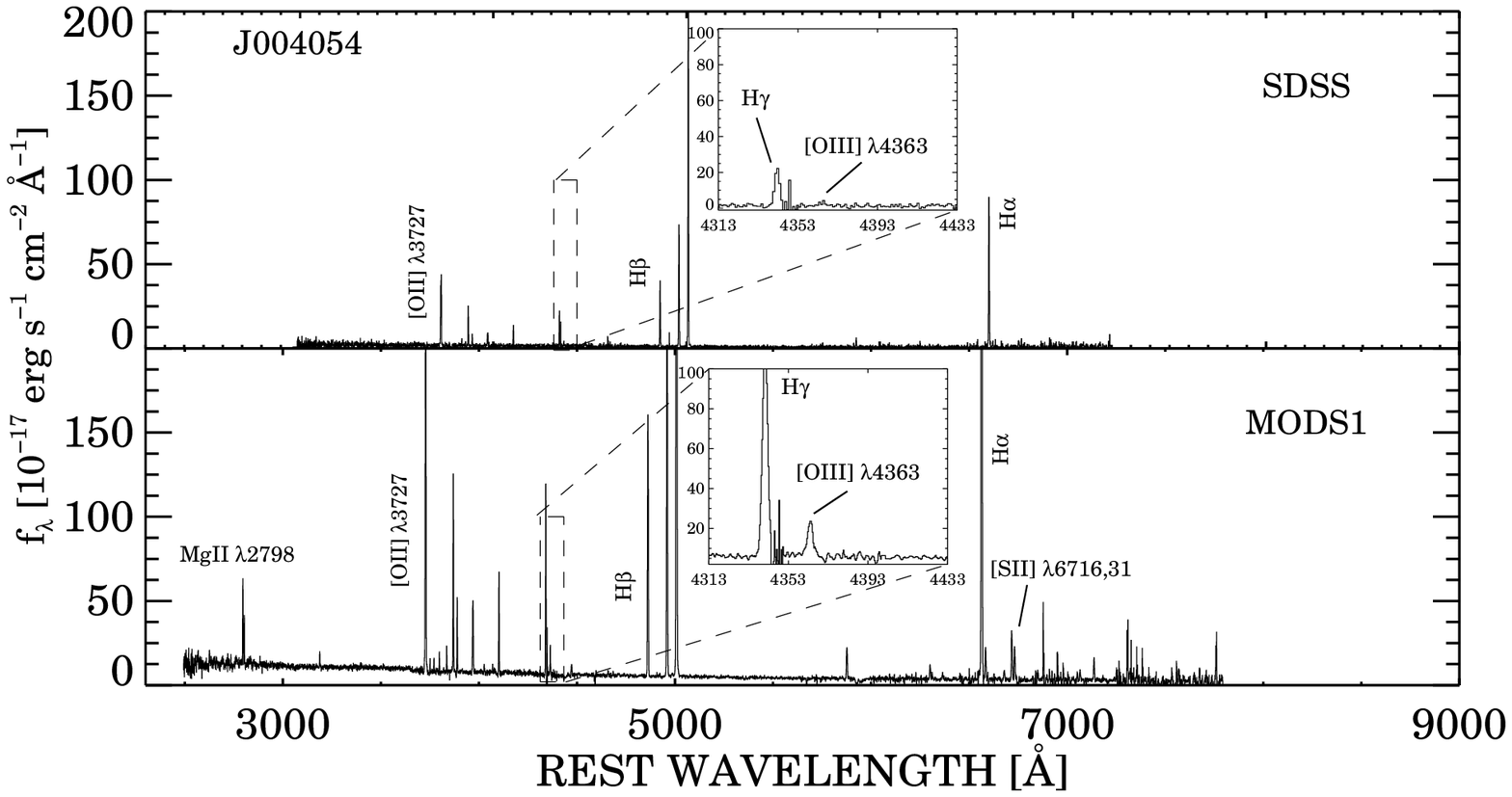}}
\end{figure*}

\begin{figure*}
\centering{\includegraphics[scale=1.,width=\textwidth,trim=0.pt 0.pt 0.pt 0.pt,clip]{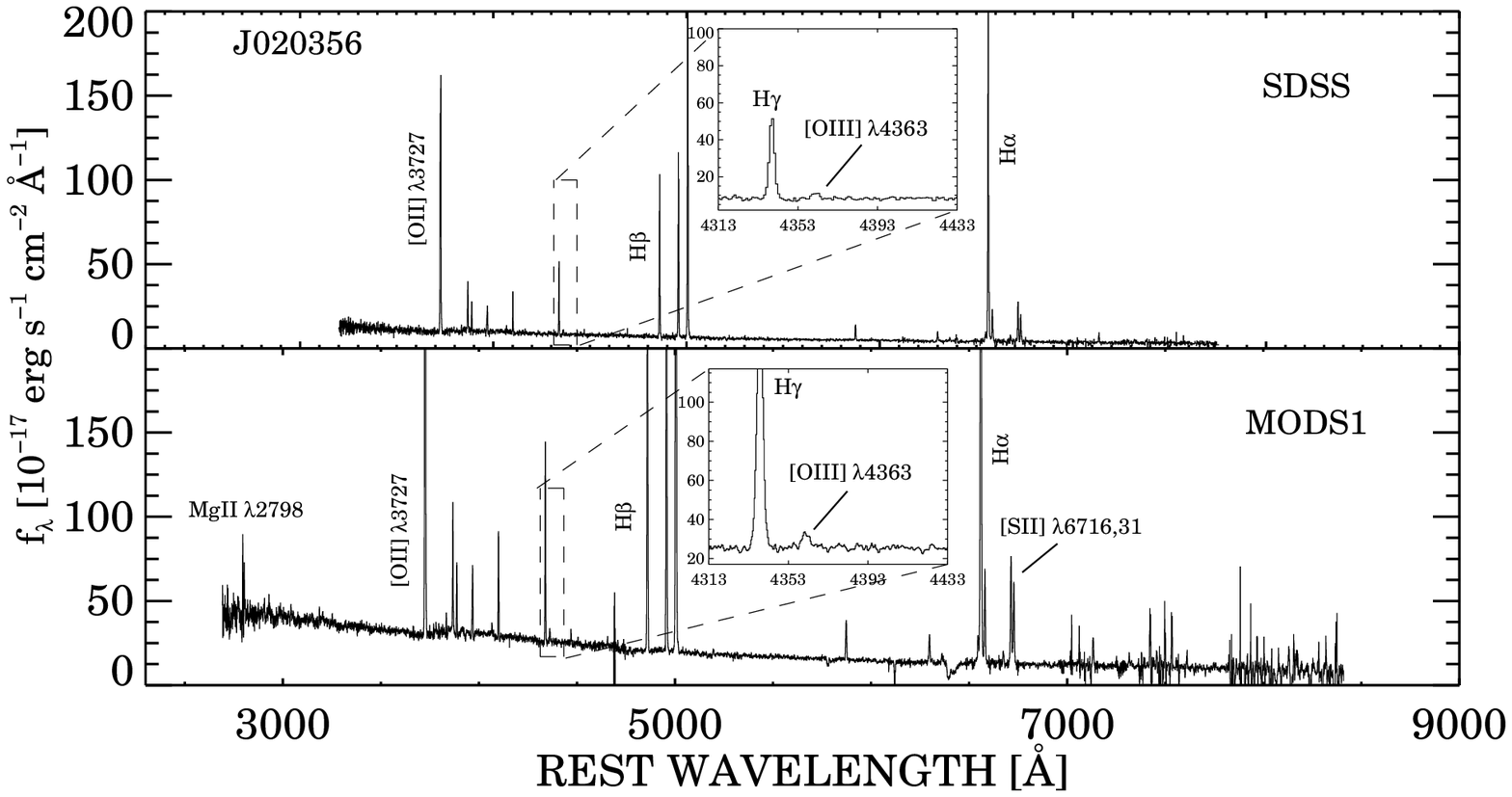}}
\end{figure*}

\begin{figure*}
\centering{\includegraphics[scale=1.,width=\textwidth,trim=0.pt 0.pt 0.pt 0.pt,clip]{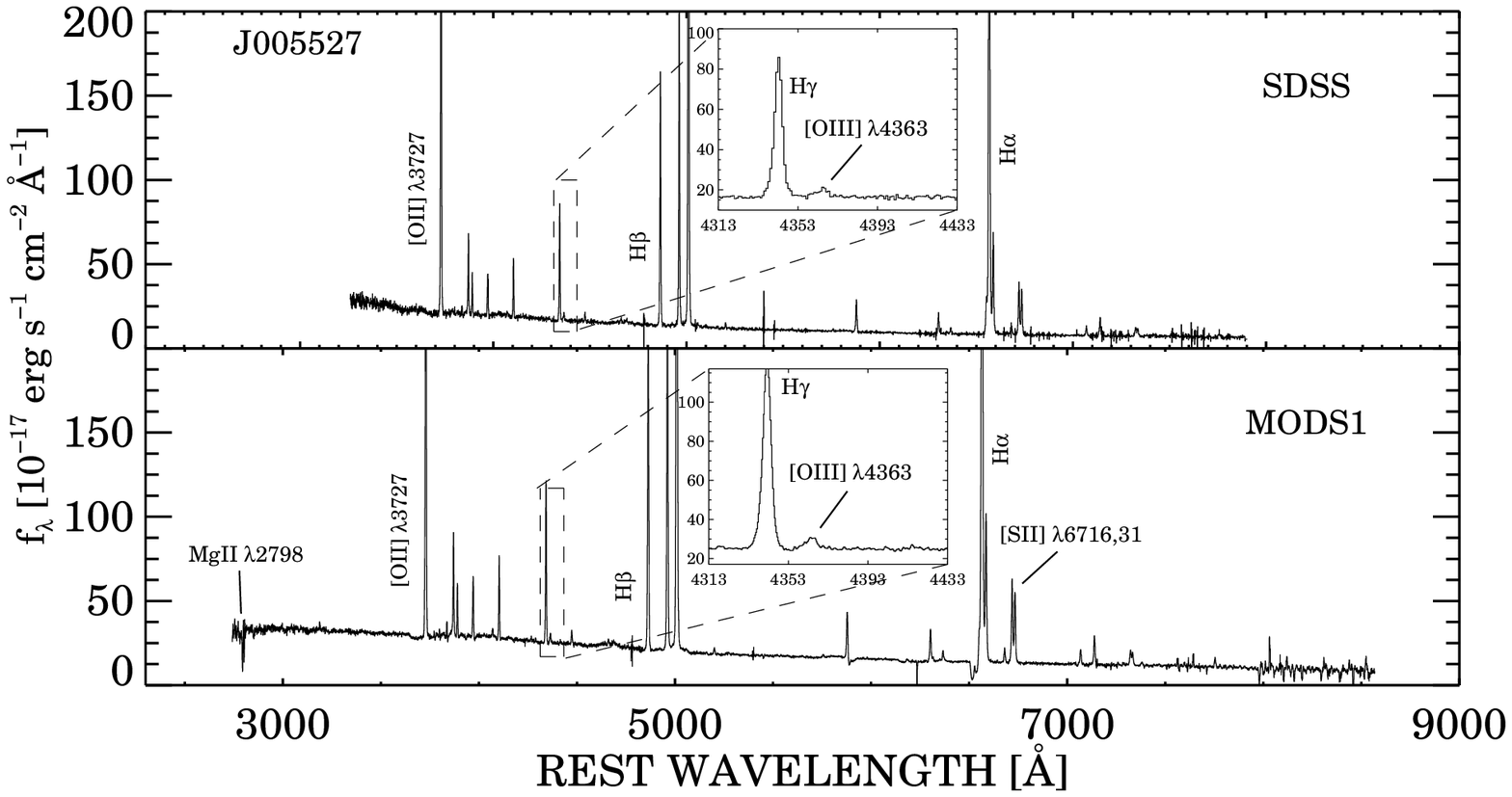}}
\caption{Comparison of our MODS1 spectra with SDSS spectra. Note the higher sensitivity and wider wavelength coverage of the MODS1 spectra. The noise spike and drop in flux seen in the inset of the J004054 MODS1 spectrum is due to the coincidental location of a strong sky line for this particular target.}
\label{fig:all_spec}
\end{figure*}

\subsection{Observing Procedures}
We observed 4 of the LBAs identified in \citet{Overzier09} using MODS1 on the LBT \citep{Pogge10} between September 2011 and January 2013. All targets were observed in longslit mode with a 1\farcs0 slit imaged onto two $3072 \times 8192$ format e2v CCDs with 15$\mu$m pixels. MODS1 uses a dichroic that splits the light into separately optimized red and blue channels at $\sim$ 5650\,\AA. The blue CCD covers a  wavelength range of $\sim$ 3200 -- 5650\,\AA, with a  400 l mm$^{-1}$ grating (spectral resolution of 2.4\,\AA), while the red CCD covers a wavelength range of $\sim$ 5650 -- 10000\,\AA, with a 250 l mm$^{-1}$ grating (spectral resolution of 3.4\,\AA).

Each target was observed with three 600s exposures for a total of 1800\,s, with the exception of J005527, which was observed with four 1200s exposures for a total of 4800\,s. The position angle of the slit approximated the parallactic angle at the midpoint of the observation so as to minimize slit losses due to differential atmospheric refraction. If the arc lamp or flat field data was not available on the night of the observation, we used the calibration data obtained within 1-2 days of our observations. Given the stability of MODS1 over the course of an observing run, this is sufficiently recent to provide accurate calibrations. We obtained bias frames and Hg(Ar), Ne, Xe, and Kr calibration lamp images, which we used for wavelength calibration. Night sky lines were used to correct for the small ($\sim$ 1\,\AA) residual flexure. Standard stars were observed with a 5x60\arcsec spectrophotometric slit mask used for flux calibration. The standard stars are from the HST Primary Calibrator list, which is composed of well observed northern-hemisphere standards from the lists of \citet{Oke90} and \citet{Bohlin95}.

Target selection was done such that priority was given to the objects from \citet{Overzier09} with the lowest oxygen abundance estimates, and hence most offset from the MZ relation, that were visible at the time of observation. Our final sample has a mean redshift $\langle z \rangle = 0.205$ with standard deviation $\sigma_z = 0.053$. 

\subsection{Data Reduction}

The basic 2D data reduction was performed in Python using the modsCCDRed suite of programs\footnote{http://www.astronomy.ohio-state.edu/MODS/Software/modsCCDRed/}. We used modsCCDRed to bias subtract, flat field, and illumination correct the raw data frames. We then coadded the frames and removed cosmic rays with L.A. Cosmic\footnote{http://www.astro.yale.edu/dokkum/lacosmic/} \citep{vanDokkum01}, taking extra care to ensure any emission features were not misidentified as cosmic rays.

We performed sky subtraction and 1D extraction using the modsIDL pipeline\footnote{http://www.astronomy.ohio-state.edu/MODS/Software/modsIDL/}. This pipeline has been developed specifically for MODS and makes use of the XIDL packages\footnote{http://www.ucolick.org/$\sim$xavier/IDL/}. Figure~\ref{fig:all_spec} shows our reduced spectra compared with spectra from the SDSS. The MODS1 spectra achieve a higher S/N than the SDSS spectra across a wider wavelength range. The inset shows a zoomed view of the metallicity sensitive $\oaur$\,\AA\ auroral emission line. For each target, the MODS1 spectra show a high significance detection of the line. In some cases, the effects of stellar absorption near the Balmer lines can be seen. Note the lack of detection of the auroral nitrogen emission line, consistent with the previously estimated low metallicity of these objects from empirical bright line methods. 

\section{Analysis}
\label{sec:analysis}
\subsection{Stellar Continuum Subtraction}
Many of the emission lines are blended with underlying stellar absorption features. To obtain accurate line flux measurements it is necessary to remove the underlying stellar component before extracting line fluxes. 

Prior to modeling the underlying stellar component of the LBAs, we correct for foreground Galactic exinction using the dust maps from \citet{Schlegel98} and the reddening law from \citet{Odonnell94} with $R_{\rm V} = 3.1$. We then shift each spectra to the rest frame using the redshifts from the SDSS and resample our spectra to 1\,\AA\ per pixel. 

We model the underlying stellar component of each target using the STARLIGHT stellar population synthesis code \citep{CidFernandes05,CidFernandes11}. STARLIGHT uses a Markov Chain Monte Carlo approach to fitting a combination of spectra from a stellar library to an observed spectrum. We adopted the stellar library from \citet{Bruzual03} and assumed a Chabrier IMF \citep{Chabrier03}. We chose base spectra that cover a wide range in both metallicity (0.0004 $\leq Z \leq$ 0.03) and age (1 Myr $\leq \tau \leq$ 10 Gyr). In general, our galaxies lack any strong stellar absorption features which could be used to place strong constraints on the underlying stellar population. However we are able to model the general shape of the continuum, as well as remove any absorption features near our metallicity-sensitive lines. In particular, the $\oaur$\,\AA\ line lies in close proximity to $\hgamma$, and so we need to take extra care to make sure the $\oaur$\,\AA\ flux in not degraded by Balmer absorption. In general, we find that our galaxies are best fit with a young ($\tau$\,$\lesssim$\,20 Myr) stellar population of roughly solar metallicity, with a velocity dispersion $\sigma~\sim~200$~km~s$^{-1}$. We regard this strictly as a qualitative assement; our targets can be equally well fit with a wide range of ages, metallicities, and kinematics. In general, we find the effects of our detailed model fit on the strengths of the stellar absorption features to be minimal. Accounting for stellar absorption, we find, on average, EW($\hbeta$(ABS)) = 2.70\,\AA, which is small compared to our mean EW($\hbeta$) $= -111$\,\AA. See Table~\ref{tab:ratio_info} for further details.

\subsection{Line Flux Measurment}

\begin{deluxetable*}{lcccc}
\tablecaption{LBA Emission Line Intensities}
\tablehead{\colhead{Ion} & \colhead{J092600} & \colhead{J004054} & \colhead{J020356} & \colhead{J005527}}
\startdata
\textrm{[\ion{O}{2}]} $\lambda$3727 &  1.576 $\pm$  0.044 &  1.662 $\pm$  0.044 &  2.684 $\pm$  0.063 &  2.082 $\pm$  0.111\\
\textrm{\ion{He}{1}} $\lambda$3820 &  0.005 $\pm$  0.009 &  0.008 $\pm$  0.007 &  0.006 $\pm$  0.014 &  0.016 $\pm$  0.010\\
\textrm{[\ion{Ne}{3}]} $\lambda$3869 &  0.468 $\pm$  0.016 &  0.485 $\pm$  0.018 &  0.315 $\pm$  0.012 &  0.308 $\pm$  0.015\\
H$\gamma$ $\lambda$4340 &  0.480 $\pm$  0.016 &  0.472 $\pm$  0.022 &  0.481 $\pm$  0.015 &  0.468 $\pm$  0.018\\
\textrm{[\ion{O}{3}]} $\lambda$4363 &  0.076 $\pm$  0.005 &  0.073 $\pm$  0.017 &  0.025 $\pm$  0.007 &  0.022 $\pm$  0.004\\
\textrm{[\ion{He}{2}]} $\lambda$4686 &  0.015 $\pm$  0.005 &  0.008 $\pm$  0.004 &  0.004 $\pm$  0.011 &  0.016 $\pm$  0.005\\
H$\beta$ $\lambda$4861 &  1.000 $\pm$  0.037 &  1.000 $\pm$  0.044 &  1.000 $\pm$  0.033 &  1.000 $\pm$  0.056\\
\textrm{[\ion{O}{3}]} $\lambda$4959 &  1.831 $\pm$  0.060 &  1.939 $\pm$  0.076 &  1.193 $\pm$  0.038 &  1.179 $\pm$  0.053\\
\textrm{[\ion{O}{3}]} $\lambda$5007 &  5.338 $\pm$  0.169 &  5.825 $\pm$  0.220 &  3.496 $\pm$  0.104 &  3.509 $\pm$  0.141\\
H$\alpha$ $\lambda$6563 &  2.927 $\pm$  0.094 &  2.878 $\pm$  0.117 &  2.930 $\pm$  0.096 &  2.861 $\pm$  0.128\\
\textrm{[\ion{N}{2}]} $\lambda$6583 &  0.119 $\pm$  0.014 &  0.105 $\pm$  0.021 &  0.246 $\pm$  0.016 &  0.431 $\pm$  0.027\\
\textrm{[\ion{S}{2}]} $\lambda$6717 &  0.161 $\pm$  0.014 &  0.158 $\pm$  0.023 &  0.308 $\pm$  0.018 &  0.232 $\pm$  0.024\\
\textrm{[\ion{S}{2}]} $\lambda$6731 &  0.121 $\pm$  0.014 &  0.107 $\pm$  0.024 &  0.226 $\pm$  0.017 &  0.198 $\pm$  0.023\\
\textrm{[\ion{O}{2}]} $\lambda$7320 &  0.022 $\pm$  0.009 &  0.049 $\pm$  0.026 &  0.022 $\pm$  0.023 &  0.035 $\pm$  0.005\\
\textrm{[\ion{O}{2}]} $\lambda$7330 &  0.017 $\pm$  0.009 &  0.013 $\pm$  0.032 &  0.008 $\pm$  0.027 &  0.030 $\pm$  0.004\\
\cutinhead{H$\beta$ Information}
C(H$\beta$) &  0.041 $\pm$  0.024 &  0.178 $\pm$  0.034 &  0.149 $\pm$  0.019 &  0.236 $\pm$  0.041\\
EW(H$\beta$(ABS))(\AA) &  2.40 $\pm$  2.00 &  3.40 $\pm$  2.00 &  3.80 $\pm$  2.00 &  3.50 $\pm$  2.00\\
EW(H$\beta$) (\AA) &   $-118$ &   $-187$ &    $-69$ &    $-70$\\
\enddata
\tablecomments{Units are such that $\hbeta = 1$.}
\label{tab:ratio_info}
\end{deluxetable*}

We assume the that emission lines in our targets are approximately Gaussian. We fit a library of atomic lines to each spectrum using MPFIT\footnote{http://purl.com/net/mpfit} \citep{Markwardt09}, an IDL implementation of the robust non-linear least squares fitting routine MINPACK-1. We assume the lines all have the same FWHM in velocity space and allow for variation in the intensity of each line as well as a small translation in wavelength ($\lesssim$\,1\,\AA). We test the robustness of our flux measurements and find that when using other flux extraction methods (e.g. direct integration), we obtain good agreement between the two methods to within a percent or so. We normalize our fluxes to $\hbeta$ and deredden the spectra using the Balmer decrement. Our final value of $C(\hbeta)$ is an error-weighted average of the values obtained using $\halpha/\hbeta$, $\hbeta/\hgamma$, and $\halpha/\hgamma$ and their appropriate intensity ratios assuming Case B recombination. See Table~\ref{tab:ratio_info} for a list of relevent line intensities used in our subsequent analysis. We adopt the following notation: 
\begin{flalign*}
&R_2=I_{\text{[\ion{O}{2}]} \lambda \lambda 3727,\,29}/I_{\hbeta} & \\
&R_3=I_{\text{[\ion{O}{3}]} \lambda \lambda 4959,5007}/I_{\hbeta} & \\
&R_{23} = R_2+R_3 & \\
&P = R_3/R_{23} & \\
\end{flalign*}
for the principal diagnostic emission line ratios.

\subsection{Abundance Determination}

We detect $\oaur$\,\AA\ in each of the galaxies, with the minimum detection at $S/N = 8.8$ and an average $S/N = 15.9$; well above the $S/N$ typically required for obtaining an electron temperature from $\oaur$\,\AA\ \citep[e.g.][]{Croxall09}. We assume the electrons in the \ion{H}{2} region follow a Maxwell--Boltzmann equilibrium energy distribution. In the low density Boltzmann regime, $\oaurratio$ is goverened by the relative level populations of the [\ion{O}{3}] ion, and, due to the spacing of the energy levels, the relative level population is very sensitive to the electron temperature of the ionized region. Since the derived ionic abundances are a strong function of electron temperature, a measurement of $\oaurratio$ allows us to compute the total oxygen abundance directly, rather than having to rely on empirical methods. 

\begin{deluxetable*}{lccccc}
\tablecaption{Electron Temperatures \& Densities}
\tablehead{\colhead{Target} & \colhead{$T_e$(OII) [K]} & \colhead{$T_e$(OIII) [K]} & \colhead{$N_e$(SII) [cm$^{-3}$]}} \\
\startdata
J092600 &  1.189 $\pm$ 0.11843 $\times 10^4$  &  1.324 $\pm$ 0.01609 $\times 10^4$  & 100\\
J004054 &  \nodata  &  1.262 $\pm$ 0.05396 $\times 10^4$  & 100\\
J020356 &  \nodata  &  1.040 $\pm$ 0.04571 $\times 10^4$  & 100\\
J005527 &  1.390 $\pm$ 0.05420 $\times 10^4$  &  1.000 $\pm$ 0.02756 $\times 10^4$  & 312\\
\enddata
\label{tab:electron_info}
\end{deluxetable*}

The assumption of a Maxwell--Boltzmann distribution of electron energies has recently come into question. Specifically, even different temperature sensitive line ratios used to directly measure the electron temperature (e.g. [\ion{O}{3}] $\lambda$4363/$\lambda$$\lambda$4959,\,5007, [\ion{S}{2}] $\lambda$$\lambda$4069,\,76/$\lambda$$\lambda$6717,\,31, [\ion{N}{2}] $\lambda$5755/$\lambda$$\lambda$6548,\,84, and [\ion{O}{2}] $\lambda$$\lambda$7318,\,24/$\lambda$$\lambda$3726,\,9) sometimes yield inconsistent results. \citet{Nicholls12} proposed that these discrepancies could be explained if the distribution of electron energies follows a $\kappa$-distribution, rather than a Maxwell--Boltzmann distribution. However, we are only concerned with {\it relative} comparisons for a given electron temperature measurement method. Thus, we do not expect systematic effects arising from our assumed energy distribution to significantly influence our results. 

\begin{figure}
\centering{\includegraphics[scale=1.,width=0.75\textwidth,trim=0.pt 0.pt 0.pt 0.pt,clip]{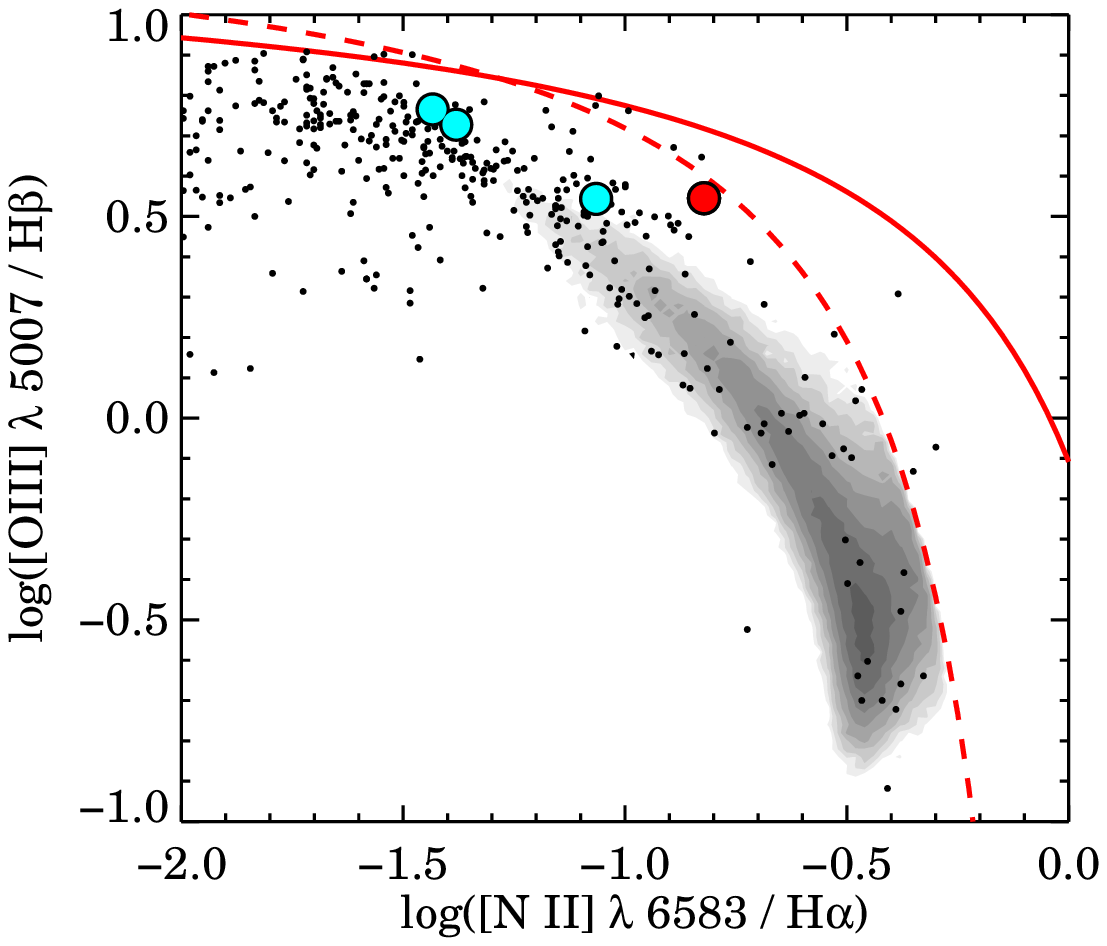}}
\caption{BPT diagram composed of star forming SDSS galaxies (gray contours), \ion{H}{2} regions from \citet{Pilyugin12} (black points), and the LBAs from our sample (cyan/red circles). The uncertainties in the line ratios for the LBAs are smaller than the plotting symbols. The dashed and solid red lines are from \citet{Kauffmann03b} and \citet{Kewley06} and denote the boundaries between star forming galaxies and AGN. The red circle represents J005527, which seems to exhibit enhanced nitrogen relative to the other LBAs.}
\label{fig:bpt}
\end{figure}

\begin{figure*}
\centering{\includegraphics[scale=1.,width=\textwidth,trim=30.pt 30.pt 20.pt 30.pt,clip]{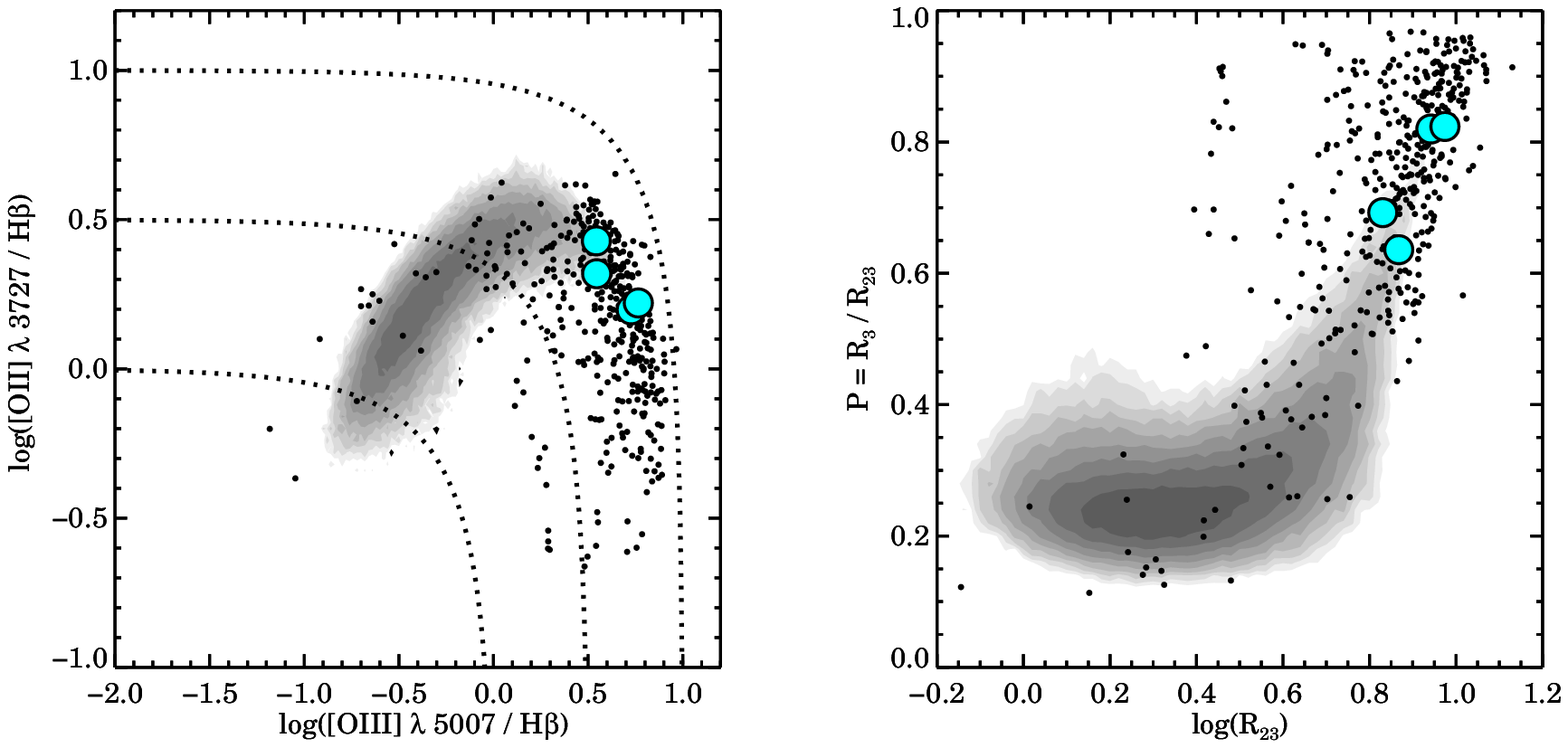}}
\caption{Excitation diagnostic plots for SDSS galaxies (gray contours), \ion{H}{2} regions from \citet{Pilyugin12} (black points), and the LBAs from our sample (cyan points). The uncertainties in the line ratios for the LBAs are smaller than the plotting symbols. The left panel shows $\thirtyseven$ as a function of $\fiveoo$. Dotted lines show constant $R_{23}$; from bottom left to top right $\log(R_{23})$ = 0., 0.5, 1.0. In general, the LBAs display higher excitation conditions more similar to \ion{H}{2} regions than the SDSS star forming galaxies. The right panel shows P $= \rthree/\rtwothree$ as a function of $\log(\rtwothree)$. Again, the LBAs occupy an excitation regime closer to that of \ion{H}{2} regions than star forming SDSS galaxies.}
\label{fig:excite}
\end{figure*}

\begin{figure*}
\centering{\includegraphics[scale=1.,width=\textwidth,trim=0.pt 150.pt 0.pt 0.pt,clip]{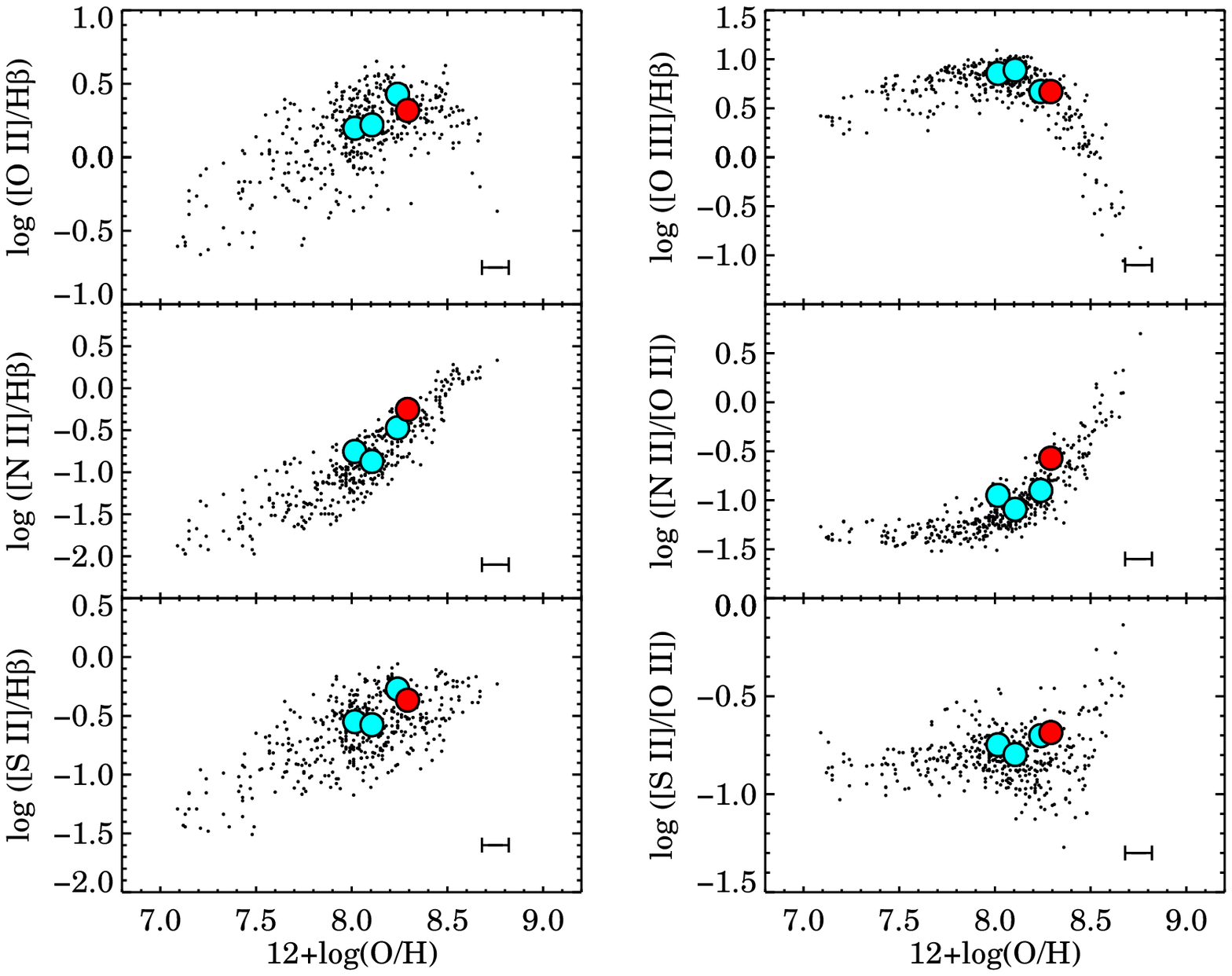}}
\caption{Diagnostic emission-line ratios as a function of oxygen abundance for our LBAs and a comparison set of local \ion{H}{2} regions from \citet{Pilyugin12}. The LBAs from our sample are shown as cyan points, with the location of J005527 marked with a red circle; the \ion{H}{2} regions with well measured metallicities from \citet{Pilyugin12} are shown as black points. The error bars in the lower right of each plot represent an uncertainty in the oxygen abundance of 0.07 dex. J005527 displays relatively stronger [\ion{N}{2}] emission compared to the other LBAs, but remains well within the parameter space occupied by the \ion{H}{2} regions. The LBA line ratios are remarkably similar to those of the ``warm'' \ion{H}{2} regions from \citet{Pilyugin10}, suggesting that the excitation conditions in LBAs are actually quite similar to that of typical \ion{H}{2} regions.}
\label{fig:pilyugin}
\end{figure*}

\citet{Osterbrock06} describes how to compute $T_e$ and $n_e$ in ionized nebulae. In this paper, we measure the electron temperature ($T_e$) and density ($n_e$) using the \verb|im_temden| IDL routines from the Moustakas code repository\footnote{https://github.com/moustakas/impro}. This set of routines uses well-known line ratios to compute the electron temperature and/or electron density of a given region. We assume a three zone ionization region, composed of a high ionization region (with $T_e = T_3 \equiv \teoiii$), an intermediate ionization region ($T_e \equiv \teariii$), and a low ionization region (with $T_e = T_2 \equiv \teoii$).  We measure $\tthree$ from $\oaurratio$ and compute $\ttwo$ using the relation

\begin{equation}
t_2 = 0.264 + 0.835t_3
\end{equation}
(where $t = T/10^4$) from \citet{Pilyugin09}. Due to the substantial noise from sky contamination and relative weakness of the [\ion{O}{2}] $\lambda \lambda$7320,\,30 lines, we don't achieve a strong detection of these lines in all of our targets. In the two cases where we are able to make a robust measurement of the lines, we find reasonable agreement between the calcuated $\ttwo$ using the relation above and the $\ttwo$ we measure from the [\ion{O}{2}] line ratios. We compute the electron density $n_e$ using the density sensitive line ratio $\sulratio$. For objects with densities less than 100~cm$^{-3}$, we adopt a value of 100~cm$^{-3}$, as lower densities are consistent with 100~cm$^{-3}$. While J005527 shows signs of slightly higher electron density, we are still well inside the low density regime for each of our targets. Table~\ref{tab:electron_info} lists the measured temperatures and densities for each LBA in our sample. Given the redshift of these objects, several [\ion{S}{3}] lines (e.g. [\ion{S}{3}] $\lambda \lambda$9069,\,9532) are unobservable with MODS1, and so we exclude sulphur from our abundance determinations.

We then take our electron temperatures and densities and use Moustakas' \verb|im_nlevel| routine to compute the relative populations and emissivities for the different ions using an n-level atom calculation. For simplicity, in the low ionization zone we adopt the reasonable canonical assumptions that $\tenii = \teoii = \ttwo$ where $\ttwo$ is the theoretical value derived from our measured $\tthree$. Similarly, for the high ionization region we assume $\teneiii = \teoiii = \tthree$. Lastly, in the intermediate ionization region, we assume $\teariii = 0.83\tthree + 0.17$ from \citet{Garnett92}. We obtain a density for each ion using

\begin{equation}
\frac{N(X^i)}{N(H^+)} = \frac{I_{\lambda_i}}{I_{\footnotesize \hbeta}} \frac{j_{\footnotesize \hbeta}}{j_{\lambda_i}}
\end{equation}

In order to compute the total abundance of a given element, we sum the observable ionic states. In the case of oxygen, we assume

\begin{equation}
{\rm \frac{O}{H} = \frac{O^0 + O^+ + O^{++}}{H^+}}
\end{equation}

We compute both total and ionic abundances for O, N, Ne, and Ar. We compute ionization correction fractions (ICFs) for N, Ne, and Ar. Adopting the ICFs from \citet{Thuan95}:

\begin{equation}
{\rm ICF(N) = \frac{O}{O^+}}
\end{equation}

\begin{equation}
{\rm ICF(Ne) = \frac{O}{O^{++}}}
\end{equation}

\begin{equation}
{\rm ICF(Ar) = \frac{Ar}{Ar^{++}}} = [0.15 + x(2.39 - 2.64x)]^{-1}
\end{equation}

where $x = {\rm O^+/O}$. The abundance estimates (and associated uncertainties) for each object is presented in Table~\ref{tab:abund_info}.

\section{Results}
\label{sec:results}
\subsection{Excitation}
In Figure~\ref{fig:bpt} we show the standard diagnostic Baldwin, Phillips, Terlevich (BPT) diagram \citep{Baldwin81} used to distiguish between ionization regions heated primarily by star-formation and regions heated primarily by AGN. Star forming SDSS galaxies\footnote{Available at http://www.mpa-garching.mpg.de/SDSS/DR7/} from the MPA-JHU catalog are shown as gray contours, \ion{H}{2} regions\footnote{\noindent Available at http://vizier.cfa.harvard.edu/viz-bin/VizieR?-source=J/MNRAS/424/2316} from \citet{Pilyugin12} are shown as black points, and our LBAs are shown as the large cyan dots. The dashed and solid red lines are from \citet{Kauffmann03b} and \citet{Kewley06} and denote the boundaries between star forming galaxies and AGN.  Our results are consistent with those presented in previous studies; the LBAs fall squarely in the star-formation dominated region of the BPT diagram. We find that J005527 (red circle) is offset to the right in Figure~\ref{fig:bpt} relative to our other LBAs and the \ion{H}{2} regions from \citet{Pilyugin12}, indicative of enhanced [\ion{N}{2}] emission. 

Figure~\ref{fig:excite} shows the excitation conditions of our LBAs based on the relative oxygen line ratios. We plot the star forming SDSS galaxies (gray contours), the \ion{H}{2} regions from \citet{Pilyugin12} (black dots), and our 4 LBAs (cyan dots). The left panel compares the relative [\ion{O}{2}] and [\ion{O}{3}] line ratios. The dotted lines show constant $R_{23}$; from bottom left to top right $\log(R_{23})$ = 0., 0.5, 1.0. As seen in Figure~\ref{fig:bpt}, the LBAs and \ion{H}{2} regions from \citet{Pilyugin12} have higher [\ion{O}{3}] emission compared to the SDSS galaxies. The right panel shows P $= R_3/R_{23}$ as a function of $\log(\rtwothree)$. Again, the LBAs display excitation conditions which are typical of \ion{H}{2} regions, but quite unusual for star forming SDSS galaxies.

In Figure~\ref{fig:pilyugin} we show line diagnostic diagrams from \citet{Pilyugin10}. The LBAs display line ratios that are remarkably similar to ``warm'' \ion{H}{2} regions. Again, the red circle marks the location of J005527, which shows signs of enhanced nitrogen relative to the other LBAs, even though it still remains well within the parameter space occupied by \ion{H}{2} regions.

\begin{figure}
\centering{\includegraphics[scale=1.,width=.75\textwidth,trim=0.pt 0.pt 230.pt 30.pt,clip]{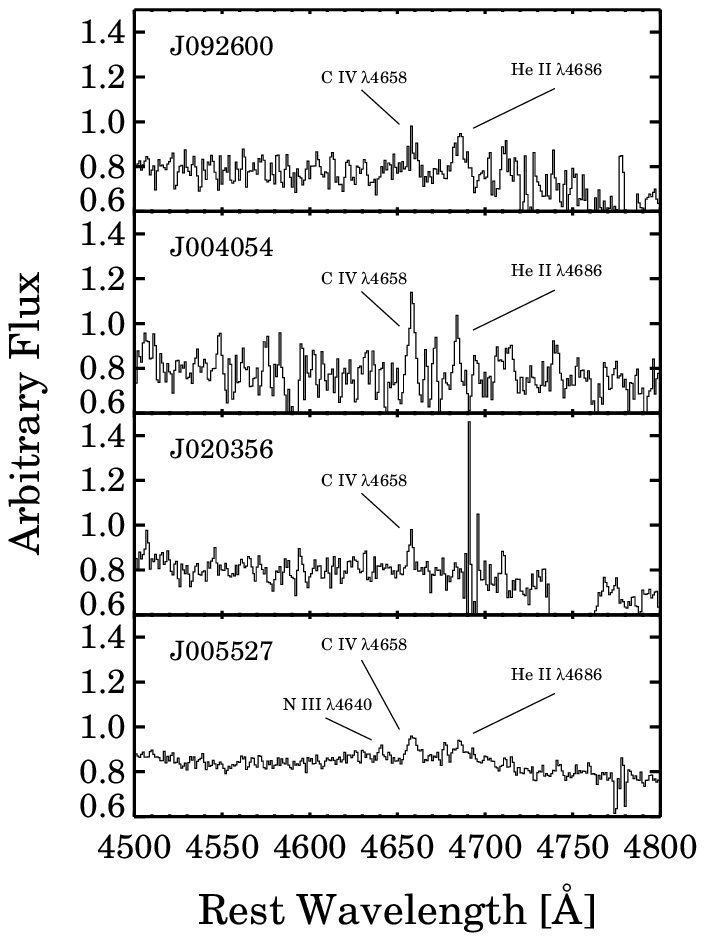}}
\caption{Wolf-Rayet features identified near 4660\,\AA\ . The most prominent features are the \ion{C}{4} line at 4658\,\AA\ and \ion{He}{2} at 4686\,\AA. J005527 displays a broad bump from 4600 -- 4700\,\AA\ as well as \ion{N}{3} emission, both of which are characteristic of a Wolf-Rayet galaxy. The noise spike and drop in flux seen in the J020356 panel is due to the coincidental location of the dichroic cutoff for this particular target.}
\label{fig:wr}
\end{figure}

Figure~\ref{fig:wr} shows the region around 4660\,\AA, where we detect characteristic Wolf-Rayet features \citep[e.g.][]{Bresolin04,Brinchmann08} in each of our 4 targets. The most common features are \ion{C}{4} $\lambda$4658 \AA, and \ion{He}{2} $\lambda$4686 \AA. J005527 displays the strongest Wolf-Rayet signatures, specifically \ion{N}{3} $\lambda$4640\,\AA\ and a broad bump from $\sim$4600--4700 \AA, in addition to the \ion{C}{4} and \ion{He}{2} emission lines. These high ionization features are associated with very young stellar populations, as they are typically visible for only a few million years following an episode of significant star formation.

\subsection{Oxygen Abundances}
Our oxygen abundances are presented in Table~\ref{tab:strong_lines}. We are able to reproduce the oxygen abundances from \citet{Overzier09} to within a few percent using the PP04 O3N2 method on the SDSS spectra. We have included the original N2 and O3N2 calibrations from PP04 as well as newer CALIFA-$T_e$ calibrations from \citet{Marino13} that use a larger sample of \ion{H}{2} regions with direct oxygen abundances. The calibrations from \citet{Marino13} are generally shallower than those presented in PP04, but in the abundance range we are concerned with, the CALIFA and PP04 calibrations produce nearly identical results. 

Due to the location of the LBAs in the transition zone of the $R_{23}$ index, the popular theoretical strong-line calibrations are rather insensitive to the oxygen abundance of these objects \citep[e.g.][]{Pilyugin05,Pena12}. Furthermore, the numerous theoretical calibrations are known to systematically deviate from each other \citep[see][]{Kewley08}. For these reasons, we consider here only the empirical strong-line methods, which are defined almost entirely by \ion{H}{2} regions with direct abundance determinations.

\begin{deluxetable*}{lccccc}
\tablecaption{Derived Abundances}
\tablehead{\colhead{Parameter} & \phantom{x} & \colhead{J092600} & \colhead{J004054} & \colhead{J020356} & \colhead{J005527}}
\startdata
${\rm O^0/H^+} (\times 10^5)$ & \phantom{x} &   0.302 $\pm$   0.025 &   0.357 $\pm$   0.067 &    \nodata &   1.162 $\pm$   0.111\\
${\rm O^+/H^+} (\times 10^5)$ & \phantom{x} &   1.787 $\pm$   0.035 &   2.155 $\pm$   0.040 &   6.042 $\pm$   0.100 &   5.588 $\pm$   0.187\\
${\rm O^{++}/H^+} (\times 10^5)$ & \phantom{x} &   8.283 $\pm$   0.176 &  10.234 $\pm$   0.277 &  11.309 $\pm$   0.245 &  12.829 $\pm$   0.380\\
\vspace{3pt} 12+log(O/H) & \phantom{x} &  8.02 $\pm$  0.06 &  8.11 $\pm$  0.08 &  8.24 $\pm$  0.05 &  8.29 $\pm$  0.08\\
${\rm N/H^+} (\times 10^5)$ & \phantom{x} &    \nodata &    \nodata &    \nodata &    \nodata\\
${\rm N^+/H^+} (\times 10^5)$ & \phantom{x} &   0.109 $\pm$   0.012 &   0.101 $\pm$   0.027 &   0.342 $\pm$   0.023 &   0.645 $\pm$   0.041\\
\phantom{x} & ICF N &   5.805 $\pm$  0.152 &   5.839 $\pm$  0.170 &   2.872 $\pm$  0.065 &   3.504 $\pm$  0.141\\
12+log(N/H) & \phantom{x} &  6.80 $\pm$  0.05 &  6.77 $\pm$  0.09 &  6.99 $\pm$  0.03 &  7.35 $\pm$  0.03\\
\vspace{3pt}log(N/O) & \phantom{x} & -1.22 $\pm$  0.26 & -1.33 $\pm$  0.47 & -1.25 $\pm$  0.16 & -0.94 $\pm$  0.17\\
${\rm Ne^{++}/H^+} (\times 10^5)$ & \phantom{x} &   1.719 $\pm$   0.058 &   2.063 $\pm$   0.068 &   2.748 $\pm$   0.107 &   3.158 $\pm$   0.154\\
\phantom{x} & ICF Ne &   1.252 $\pm$  0.035 &   1.245 $\pm$  0.044 &   1.534 $\pm$  0.041 &   1.526 $\pm$  0.057\\
12+log(Ne/H) & \phantom{x} &  7.33 $\pm$  0.14 &  7.41 $\pm$  0.15 &  7.62 $\pm$  0.16 &  7.68 $\pm$  0.20\\
\vspace{3pt}log(Ne/O) & \phantom{x} & -0.68 $\pm$  0.09 & -0.70 $\pm$  0.10 & -0.61 $\pm$  0.10 & -0.61 $\pm$  0.13\\
${\rm Ar^{++}/H^+} (\times 10^5)$ & \phantom{x} &   0.032 $\pm$   0.004 &   0.039 $\pm$   0.010 &    \nodata &   0.061 $\pm$   0.004\\
\phantom{x} & ICF Ar &   2.069 $\pm$  0.207 &   2.089 $\pm$  0.209 &   \nodata &   1.621 $\pm$  0.162\\
12+log(Ar/H) & \phantom{x} &  5.82 $\pm$  0.42 &  5.91 $\pm$  0.73 &  \nodata &  5.99 $\pm$  0.32\\
log(Ar/O) & \phantom{x} & -2.19 $\pm$  0.38 & -2.20 $\pm$  0.66 &  \nodata & -2.30 $\pm$  0.28\\
\enddata
\label{tab:abund_info}
\end{deluxetable*}

Given the scatter in the emprirical calibrations of $\sim$\,0.3~dex and the uncertainties in our direct abundances, the two results are roughly consistent in the sense that both indicate that these LBAs have low oxygen abundances for their mass relative to the MZ relation. The N2 estimates generally give higher O/H values than both the O3N2 estimates and our direct abundances (see Table~\ref{tab:strong_lines}). This is qualitatively consistent with \citet{Kewley08}, who show that the N2 method tends to yield O/H values that are slightly higher than the O3N2 method at low oxygen abundances (12+$\log$(O/H)$\lesssim 8.2$). However, with four objects in our sample this is not statistically significant, and so we refrain from drawing any inferences regarding the systematic effects between strong-line calibrations and direct abundances in LBAs. Importantly, the LBAs remain significantly below the locus of SDSS galaxies regardless of the abundance method used. With this in mind, and for the sake of simplicity and consistency with previous studies, we restrict our discussion of strong-line estimates to the N2 empirical calibration from PP04.

\begin{deluxetable*}{lcccc}
\tablecaption{Direct and Strong-Line Oxygen Abundances}
\tablehead{\colhead{Method} & \colhead{J092600}  & \colhead{J004054} & \colhead{J020356} & \colhead{J005527}}
\startdata
Direct (this work) &  8.02 $\pm$  0.06 &  8.11 $\pm$  0.08 &  8.24 $\pm$  0.05 &  8.29 $\pm$  0.08\\
\cutinhead{Strong Line Estimates -- This Work}
PP04 N2 &  8.127 &  8.107 &  8.251 &  8.375\\
PP04 O3N2 &  8.056 &  8.026 &  8.215 &  8.293\\
CALIFA N2 &  8.105 &  8.081 &  8.251 &  8.363\\
CALIFA O3N2 &  8.082 &  8.062 &  8.189 &  8.240\\
\cutinhead{Overzier Estimates}
PP04 O3N2 & 8.09 & 8.03 & 8.21 & 8.28\\
\enddata
\tablecomments{The scatter in these empirical calibrations is $\sim$\,0.3 in 12+log(O/H); see \citet{Pettini04} and \citet{Marino13} for details.}
\label{tab:strong_lines}
\end{deluxetable*}

\begin{figure}
\psfrag{X}[c][][1.]{$\odot$}
\centering{\includegraphics[scale=1.,width=\textwidth,trim=30.pt 30.pt 0.pt 30.pt,clip]{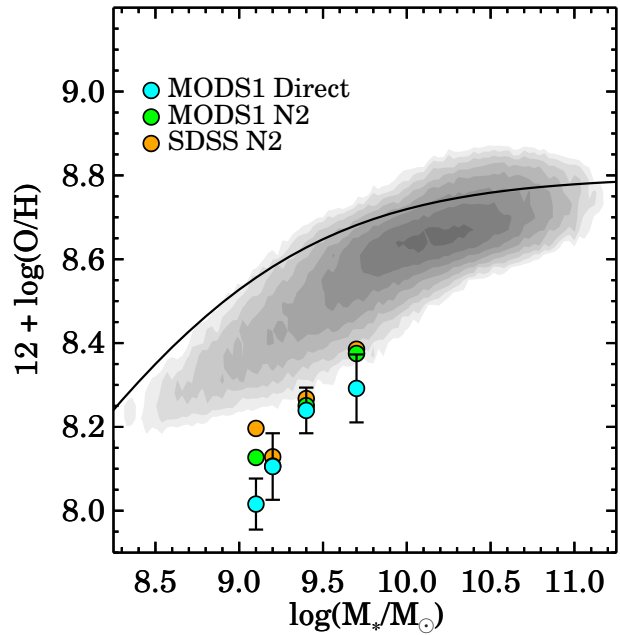}}
\caption{Gas phase oxygen abundance as a function of mass for various abundance diagnostics. The cyan points show our direct method oxygen abundance for the 4 LBAs in our sample, with masses taken from \citet{Overzier09}. The thick black line denotes the logarithmic best fit to the stacked SDSS direct oxygen abundances from \citet{Andrews13}. The orange and green points show the PP04 N2 oxygen abundance estimates from the SDSS and MODS1 spectra respectively. The gray contours are the star forming SDSS galaxies from the MPA-JHU catalogue with oxygen abundances estimated from the PP04 N2 calibration. Regardless of the diagnostic used, the LBAs display low oxygen abundances for their mass relative to the MZ relation.}
\label{fig:mm}
\end{figure}

Figure~\ref{fig:mm} shows the oxygen abundances of our LBAs relative to the SDSS galaxies. In order to minimize systematic effects, it is necessary to only compare the oxygen abundances of objects when using the same diagnostic (e.g. direct method abundance of SDSS galaxies versus direct method abundance of LBAs). Our direct method measurements of the LBAs are shown as cyan dots. Due to the difficulty of measuring the auroral lines, we do not have direct method abundances for a statistically significant number of individual SDSS galaxies. We do however have the best fit relation to the direct method oxygen abundance of stacked SDSS spectra from \citet{Andrews13}, which we have included in Figure~\ref{fig:mm} as the thick black line. 

The SDSS galaxies (gray contours) have masses from MPA-JHU catalogue (see \citet{Kauffmann03a,Salim07}) and have had the oxygen abundances estimated using the N2 calibration from PP04. The LBA N2 estimates of oxygen abundance are shown as orange and green dots for the SDSS and MODS1 data respectively; the offset from the locus of individual SDSS contours is clear and consistent with \citet{Overzier10}. While the relation from \citet{Andrews13} systematically deviates from the N2 oxygen abundance estimates (as expected), our direct method oxygen abundances of the LBAs still fall well below that of the stacked SDSS spectra. Thus it appears that systematic effects in the strong-line calibrations cannot explain the offset of the LBAs from the MZ relation; these LBAs have low oxygen abundance given their mass.

\section{Discussion}
\label{sec:discussion}

\subsection{Excitation Conditions of LBAs}

Empirical abundance calibrations are typically based on samples of \ion{H}{2} regions with well-determined electron temperatures and thus directly measured abundances. However, the direct method is subject to a number of observational biases. For instance, metal poor objects generally have higher electron temperatures, brighter auroral lines, and more easily determined abundances. This results in a preferential selection of low metallicity objects in empirical calibrations. Furthermore, most large scale surveys of emission line galaxies, like the SDSS, are composed of predominantly low-excitation galaxies relative to the \ion{H}{2} regions on which the empirical calibrations are based (see Figure~\ref{fig:excite}). \citet{Moustakas10} cautions against haphazardly extrapolating empirical relations to lower excitation regimes occupied by the majority of galaxies in large surveys, as doing so could result in erroneous abundance determinations.  

Looking at Figure~\ref{fig:pilyugin}, the emission line flux ratios of all 4 of our LBAs are remarkably similar to those of the ``warm'' \ion{H}{2} ratios from \citet{Pilyugin10}. Additionally, even though J005527 displays fairly strong [\ion{N}{2}] emission compared to the other LBAs in our sample, it remains in an excitation regime fairly typical of local \ion{H}{2} regions. This supports the idea that locally determined empirical calibrations ought to return reasonable abundance estimates for LBAs. This is a crucial step in justifying the application of locally calibrated empirical relations to LBAs and LBGs. However, caution must still be exercised when extrapolating these relations to high redshifts, as recent work has suggested that the LBG population as a whole evolves rapidly with redshift \citep{Stanway14}.

It is readily apparent that LBAs are undergoing an episode of significant star formation. Within a few million years of the initial burst, large numbers of Wolf-Rayet stars, supernova remnants, and other extremely hot objects could conspire to produce an abnormally hard ionization spectrum. If we suppose that the gas surrounding these hot objects was subject to a harder ionization spectrum than what is observed in local \ion{H}{2} regions, we would expect to see relatively enhanced ionic emission features. High energy photons have a smaller cross section for interaction. As a result, these high energy photons have a longer mean free path, resulting in larger partially ionized regions. In general, this results in enhanced ionic emission. In the case of ionized nitrogen, an anomalously hard ionization spectrum would produce enhanced [\ion{N}{2}] emission and result in systematically {\it high} abundance estimates when using a locally calibrated empirical relation.

\citet{Berg11} showed that enhanced nitrogen abundance (relative to oxygen) could also bias strong-line estimates towards high oxygen abundance. However, the excitation conditions of our LBAs are quite different from what is expected in the evolved Wolf-Rayet galaxies from \citet{Berg11}. Furthermore, our LBAs do not show abnormally high [\ion{N}{2}]/[\ion{O}{2}] ratios. If the high N2 method estimates observed in some of our targets were the result of an abnormally hard ionization spectrum, we would not expect the O3N2 method to yield a high abundance estimate, as such a radiation field would have a similar effect on both N$^{+}$ and O$^{++}$ emission.

\subsection{LBAs and the Fundamental Metallicity Relation}

\begin{figure}
\psfrag{X}[c][][1.]{$\odot$}
\centering{\includegraphics[scale=1.,width=\textwidth,trim=30.pt 30.pt 0.pt 30.pt,clip]{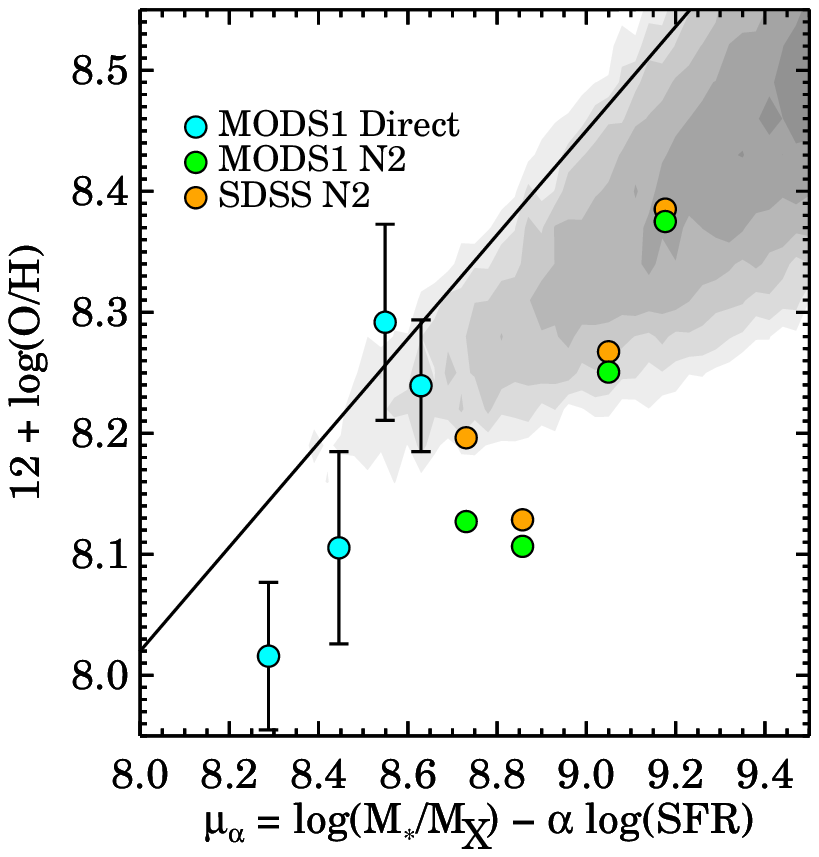}}
\caption{Gas phase oxygen abundance as a function of $\mu_{\alpha}$, where $\alpha$ is chosen depending on the abundance diagnostic. We adopt the appropriate $\alpha$ from \citet{Andrews13}; for the direct method we use $\alpha = 0.66$, and for the N2 method we use $\alpha = 0.30$. The color scheme is the same as that from~\ref{fig:mm}. Masses and $\halpha + 24 \mu m$ star formation rates for the LBAs are taken from \citet{Overzier09}; star formation rates for the SDSS galaxies are from the MPA-JHU catalogue. The thick black line denotes the linear best fit to the FMR of direct method stacked SDSS galaxies from \citet{Andrews13}. The incorporation of star formation rate reduces the degree to which the LBAs are offset from typical SDSS star forming galaxies.}
\label{fig:fmr}
\end{figure}

LBAs have masses typical of entire galaxies and thus it is interesting to note that their excitation conditions fall in a region that is sparsely populated by the SDSS star forming galaxies, but well occupied by the \ion{H}{2} regions from \citet{Pilyugin12}. It seems that the LBAs more closely resemble giant \ion{H}{2} regions in terms of photoionization conditions than typical SDSS galaxies. What is the primary physical process causing the LBAs to be so significantly offset from SDSS galaxies? 

The defining characteristic of LBAs is their compact UV emission arising from recent star formation, and indeed, the typical star formation rate for an LBA is an order of magnitude higher than a typical SDSS galaxy \citep{Overzier09}. Such high rates of star formation imply a recent replenishment of star forming material in the LBA systems. This influx of gas could be due to a tidal interaction \citep[e.g.][]{Peeples09}, or perhaps a recent merger. Generally these events will result in dilution of the interstellar gas and a corresponding reduction in the observed metallicity. 

\citet{Peeples09} compute the expected dilution which might result from the funneling of gas from large galactocentric radius into the center of a galaxy. They find that for a galaxy with a reasonable metallicity gradient and gas surface density profile, gas flowing inward from within 20 kpc would result in a metallicity dilution $\Delta$(O/H) $= -0.5$~dex, which is remarkably close to the MZ relation offsets observed for their morphologically disturbed galaxies. The LBAs in our sample also tend to be offset from the SDSS MZ relation of \citet{Andrews13} by a comparable amount, suggesting that LBAs may have recently experienced an inflow of unenriched gas. This is consistent with the disturbed morphology of LBAs seen in the HST images from \citet{Overzier09,Overzier10} and the dispersion dominated kinematics (e.g. $\sigma/v >1$) presented in \citet{Goncalves10}.

A global inflow of star forming material could also result in an intense burst of star formation \citep{Rupke08,Kewley10}. It is often assumed that the timescale for the enrichment of the interstellar medium (ISM) is short compared to the overall galaxy evolution timescale. However, the photoionization conditions of LBAs indicate that we are observing the actual burst of star formation take place. The LBAs have not had time to convert the infalling gas into stars and have those stars chemically enrich their ISM. If the burst of star formation in LBAs is indeed being powered by relatively low metallicity gas, we would expect that their residuals from the MZ relation correlate with star formation rate. 

The Fundamental Metallicity Relation \citep[FMR;][]{Mannucci10, Andrews13} parametrizes the correlation of residuals from the MZ relation and star formation rate by introducing the parameter $\mu_{\alpha}$ such that $\mu_{\alpha} = \log{M_*} - \alpha\log({\rm SFR})$. The sample of galaxies compiled by \citet{Mannucci10} consists of $>140000$ SDSS galaxies at $z \sim 0$, 182 objects from $0.5 < z < 2.5$, the 91 galaxies from \citet{Erb06}, and an additional 16 galaxies from $3 < z < 4$. Typically, $\alpha$ is chosen such that it minimizes the scatter in the resulting FMR. \citet{Mannucci10} find $\alpha = 0.32$ for their sample of galaxies, and no evolution of the FMR up to $z=2.5$. 

The exact value of $\alpha$ is quite sensitive to the abundance diagnostic used. \citet{Yates12} found $\alpha=0.19$ when using the oxygen abundances from \citet{Tremonti04}, and \citet{Andrews13} found $\alpha=0.66$ when using direct method abundances, both of which are significantly different from the $\alpha=0.32$ presented in \citet{Mannucci10}. Additionally, the determination of $\alpha$ merely minimizes the scatter for a given abundance diagnostic; two strong-line calibrations will generally not share the same FMR.

Determining where an object sits on the FMR requires knowledge of the SFR in addition to the mass and metallicity of the object. \citet{Overzier09} adopts SFR calibrations from \citet{Calzetti09} and computes various SFRs using $\halpha$, $\halpha+24\mu$m, and FUV luminosities. The $\halpha$ flux is associated with only the most recent star formation activity, whereas the FUV calibration is sensitive to the integrated star formation activity over the previous 1 Gyr. The appearance of the LBAs is dominated by the current burst of star formation activity, so we adopt the $\halpha+24\mu$m SFRs \citep{Kennicutt07,Calzetti07}. 

The $\halpha+24\mu$m calibration is not without its own systematic effects. For example, an AGN could preferentially heat dust and result in 24$\mu$m flux above that which would arise from star formation alone. However, none of the \citet{Overzier09} LBAs appear to host a Type 1 (unobscured) AGN. While the presence of Type 2 (obscured) AGN is not ruled out, it seems unlikely given where these LBAs lie on the diagnostic diagrams \citep{Overzier09}. If an AGN were present, it is likely to only have a very small effect. 

Figure~\ref{fig:fmr} shows the oxygen abundances of the SDSS star forming galaxies and the LBAs as a function of $\mu_{\alpha}$, where we have adopted the values of $\alpha = 0.30$ and $\alpha = 0.66$ corresponding to the N2 index and direct method respectively from \citet{Andrews13}. The cyan dots show our LBAs with direct method oxygen abundances; the thick black line shows the linear fit to the FMR for stacked SDSS spectra from \citet{Andrews13}. The N2 estimates of oxygen abundance for our LBAs are shown as orange and green dots for the SDSS and MODS1 data respectively. The gray contours are the star forming SDSS galaxies from the MPA-JHU catalogue (see \citet{Kauffmann03a} and \citet{Salim07} for mass determinations and \citet{Brinchmann04} for star formation rate determinations). The oxygen abundances of the SDSS galaxies are estimated from the PP04 N2 calibration. We see that plotting oxygen abundance as a function of both mass and star formation rate does indeed reduce the scatter between the LBAs and SDSS data for a given abundance estimation method (direct or emprical). 

It is important to keep in mind that a comparison of where the direct abundance measurements fall on the FMR relative to the contoured SDSS data is meaningless, since we do not expect the different abundance diagnostics to produce consistent results. However, the fact that incorporating SFR drastically reduces the scatter for a given abundance diagnostic suggests that the high SFR, low oxygen abundance, and disturbed morphology of these LBAs could be explained by a recent inflow of relatively unenriched gas and is consistent with the existence of a FMR that the LBAs appear to follow.

\section{Summary}
\label{sec:summary}
It is believed that LBAs ($z \sim 0.2$) are true analogs of LBGs ($z \gtrsim 3$), and thus laboratories for studying one of the dominant modes of star formation in the early universe in exquisite detail. Before applying locally calibrated empirical relations to these LBAs, it is important to investigate whether or not the locally calibrated empirical relations based on single, bright \ion{H}{2} regions in normal galaies still hold for the physical conditions present in LBAs. 

The empirically-derived oxygen abundances of LBAs show them to be metal deficient for their mass, falling $\gtrsim 0.2$~dex below the MZ relation defined by local star forming galaxies. We have presented direct abundance measurements of 4 LBAs using MODS1 on the LBT to detect $\oaur$\,\AA\ in each target, allowing for a direct measurement of the electron temperature and thus a robust determination of the gas phase oxygen abundance. We have shown that:

\begin{itemize}
\item LBAs display excitation conditions that are unusual for SDSS galaxies, but are quite typical of \ion{H}{2} regions from \citet{Pilyugin12}.
\item The N2 empirical calibration is generally valid for the LBAs presented here. Objects with particularly hard ionizing spectra may have biased strong-line abundance estimates, but the effect is likely to be smaller than the scatter in the empirical calibrations. 
\item LBAs are offest from the MZ relation of local star forming glaxies in the sense that they have lower oxygen abundances for a given mass. However, when their abnormally high star formation rates are taken into account, we find that they do not appear to deviate significantly from the FMR. This, coupled with their disturbed morphologies, is consistent with an interaction driven gas inflow paradigm.
\end{itemize}

We can improve our understanding of LBAs in a statistical sense by increasing the size of the sample studied. Here we have presented observations of only 4 of the 31 LBAs in the \citet{Overzier09} sample. With instruments like MODS1, precise spectroscopic observations of LBAs are quite feasible and can be done with a modest amount of telescope time.

An increased number of LBAs with robustly determined abundances would allow us to place tighter constraints on the systematic effects between locally calibrated strong-line abundance estimates and direct method abundances. This will aid greatly in our understanding of how LBAs differ from local galaxies, and improve our understanding of the processes governing the observed trends in mass, metallicity, and star formation rate.

Lastly, it has been suggested that both the MZ relation and FMR are a consequence of the relation between gas phase oxygen abundance and stellar-to-gas mass ratio \citep[the Universal Metallicity Relation;][]{Zahid14}. They argue that once the ISM of a galaxy has been enriched to a point such that the amount of oxygen being locked up in low mass stars is comparable to the oxygen produced by massive stars, the oxygen abundance asymptotically approachs a value which is independent of redshift. If the LBAs have indeed experienced a significant inflow of gas mass relative to their stellar mass, they could potentially serve as key testing grounds for the Universal Metallicity Relation.

\acknowledgements 
We would like to thank the referee for a constructive report. This paper uses data taken with the MODS spectrographs built with funding from NSF grant AST-9987045 and the NSF Telescope System Instrumentation Program (TSIP), with additional funds from the Ohio Board of Regents and the Ohio State University Office of Research. KVC and MODS pipeline software development was supported by NSF grant AST-1108693.

We appreciate the MPA-JHU group for making their catalog publicly available, as well Leonid Pilyugin, Eva Grebel, and Lars Mattsson for making their catalog of \ion{H}{2} regions available. We also thank John Moustakas for making his routines publicly available. The STARLIGHT project is supported by the Brazilian agencies CNPq, CAPES and FAPESP and by the France-Brazil CAPES/Cofecub program.

Funding for the SDSS and SDSS-II has been provided by the Alfred P. Sloan Foundation, the Participating Institutions, the National Science Foundation, the U.S. Department of Energy, the National Aeronautics and Space Administration, the Japanese Monbukagakusho, the Max Planck Society, and the Higher Education Funding Council for England. The SDSS Web Site is http://www.sdss.org/.

The SDSS is managed by the Astrophysical Research Consortium for the Participating Institutions. The Participating Institutions are the American Museum of Natural History, Astrophysical Institute Potsdam, University of Basel, University of Cambridge, Case Western Reserve University, University of Chicago, Drexel University, Fermilab, the Institute for Advanced Study, the Japan Participation Group, Johns Hopkins University, the Joint Institute for Nuclear Astrophysics, the Kavli Institute for Particle Astrophysics and Cosmology, the Korean Scientist Group, the Chinese Academy of Sciences (LAMOST), Los Alamos National Laboratory, the Max-Planck-Institute for Astronomy (MPIA), the Max-Planck-Institute for Astrophysics (MPA), New Mexico State University, Ohio State University, University of Pittsburgh, University of Portsmouth, Princeton University, the United States Naval Observatory, and the University of Washington.

\bibliography{biblio}
\end{document}